\documentclass[conference]{IEEEtran}
\IEEEoverridecommandlockouts

\usepackage{kotex}
\usepackage{cite}
\usepackage{amsmath,amssymb,amsfonts}
\usepackage{algorithm}
\usepackage{algpseudocode}
\usepackage{graphicx}
\usepackage{xcolor}
\usepackage{comment}
\usepackage{makecell}
\usepackage{pifont}
\usepackage{tabularx}
\usepackage{booktabs}
\usepackage{url}
\usepackage{subcaption}
\usepackage{setspace}

\usepackage[
    colorlinks=true,
    linkcolor=purple,
    citecolor=purple,
    urlcolor=black
]{hyperref}

\def\BibTeX{{\rm B\kern-.05em{\sc i\kern-.025em b}\kern-.08em
    T\kern-.1667em\lower.7ex\hbox{E}\kern-.125emX}}
\begin{document}

\newcommand{\secref}[1]{\hyperref[#1]{\S~\ref*{#1}}}

\newcommand{\squishlist}{
\begin{list}{$\bullet$}
	{ \setlength{\itemsep}{0pt}      \setlength{\parsep}{-0pt}
		\setlength{\topsep}{4pt}       \setlength{\partopsep}{0pt}
		\setlength{\listparindent}{-2pt}
		\setlength{\itemindent}{-5pt}
		\setlength{\leftmargin}{1em} \setlength{\labelwidth}{0em}
		\setlength{\labelsep}{0.5em} } }

\newcommand{\squishend}{\end{list}  }

\newcommand{\tbdfixed}{{\textrm{eFDP}}}
\newcommand{\tbdfixedmonitor}{{\textrm{eFDP Monitor}}}
\newcommand{\tbdfixedplanner}{{\textrm{eFDP Planner}}}
\newcommand{\tbdfixedloader}{{\textrm{eFDP Loader}}}

\newcommand{\cp}{\textcolor{pink}}
\newcommand{\cb}{\textcolor{blue}}
\newcommand{\cred}{\textcolor{red}}
\newcommand{\cg}{\textcolor{green}}
\newcommand{\cm}{\textcolor{magenta}}
\newcommand{\cpp}{\textcolor{purple}}

\newcommand{\finding}[2]{\par\smallskip\noindent\fcolorbox{black!55}{black!5}{\parbox{\dimexpr\columnwidth-2\fboxsep-2\fboxrule\relax}{\small\textbf{Finding #1.} #2}}\par\smallskip}
\newcommand{\problem}[1]{\par\smallskip\noindent\fcolorbox{black!55}{black!5}{\parbox{\dimexpr\columnwidth-2\fboxsep-2\fboxrule\relax}{\small\textbf{Problem.} #1}}\par\smallskip}

\newcommand{\tyes}{\ding{51}}
\newcommand{\tno}{\ding{55}}
\newcommand{\tpart}{$\triangle$}
\newcommand{\tsome}{$\circ$}
\newcommand{\tvhead}[1]{\rotatebox{90}{#1}}

\newcommand{\stepA}[1]{\ding{\number\numexpr181+#1\relax}}
\newcommand{\stepB}[1]{\ding{\number\numexpr171+#1\relax}}

\title{eBPF-Driven Adaptive Lifetime-Aware Data Placement for Shared FDP SSDs}

\author{
{\rm Hajin Moon$^{1}$, Taewan Noh$^{1}$, Chaeeun Yoon$^{1}$,
Jeonghun Lee$^{1}$, Junyoung Han$^{2}$, Joo-Young Hwang$^{2}$,
Youngjae Kim$^{1}$\thanks{Corresponding author.}}\\[5pt]
$^{1}$Sogang University, Seoul, Korea \quad
$^{2}$Samsung Electronics Co.
}

\maketitle

\setstretch{0.975}
\begin{abstract}
NVMe Flexible Data Placement (FDP) reduces SSD write amplification by separating data with different lifetimes to minimize garbage collection (GC) copying.
However, on shared servers, multiple tenants generate more lifetime classes than the available placement handles. 
Existing approaches rely on static mappings or application-specific knowledge, failing to capture runtime lifetimes and adapt to changing workloads.
The challenge is to identify which data die together and determine which classes should be separated to maximize GC savings under a limited handle budget.
We present \tbdfixed{}, an eBPF-based system that automatically learns data lifetimes and manages FDP placement for unmodified applications on a stock Linux kernel.
By tracking writes and invalidations, \tbdfixed{} identifies lifetime classes and estimates the GC copying saved by separating them.
It first assigns each tenant a handle, then allocates the remaining handles to the separations offering the greatest savings.
It also re-plans placement when active tenants change.
With four database tenants on an emulated FDP SSD, \tbdfixed{} eliminates twice as much excess write amplification as tenant-level isolation and improves every writing tenant's throughput by 19--23\%.
On an enterprise FDP SSD, given RocksDB's levels, it matches a manually optimized placement using three handles instead of five.
After a tenant switch, \tbdfixed{} re-plans within seven minutes and maintains 25\% lower write amplification than the original layout even after an hour.
\end{abstract}


\section{Introduction}
\label{sec:intro}

\begin{figure*}[!t]
    \centering
    \begin{subfigure}[b]{0.6\textwidth}
        \centering
        \includegraphics[height=3.0cm, width=\textwidth, keepaspectratio]{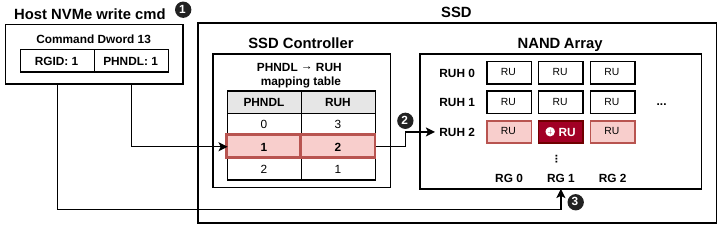}
        \caption{FDP inside the SSD}
    \end{subfigure}
    \hspace{0.02\textwidth}
    \begin{subfigure}[b]{0.3\textwidth}
        \centering
        \includegraphics[height=3.2cm, width=\textwidth, keepaspectratio]{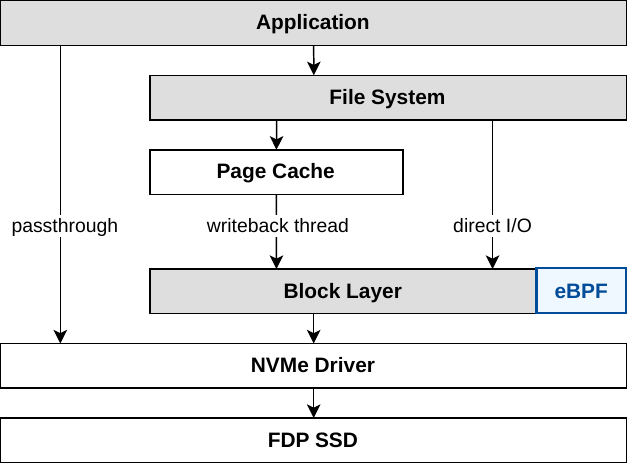}
        \caption{The Linux write path}
    \end{subfigure}
\caption{
How a write's handle is set inside the SSD (a), and where on the Linux write path (b).
(a)\,The PID holds a placement handle (PHNDL) and a reclaim group (RGID) (\ding{182}).
The controller maps PHNDL to a reclaim unit handle (RUH), our handle (\ding{183}), and RGID to a group (\ding{184}).
The data land in that handle's open RU (\ding{185}).
(b)\,The routes that can set a write's stream today (gray), and where \tbdfixed{} sets it (blue).
}
\vspace{-12pt}
\label{fig:fdp}
\end{figure*}

How an SSD groups data decides how much \emph{garbage collection} (GC) copies.
Flash cannot be overwritten in place, so an SSD writes every update to a fresh page and leaves the old copy stale, or \emph{dead}.
GC reclaims dead pages only by erasing a whole block, after copying out every page still alive~\cite{agrawal2008flash}.
These copies raise the \emph{write amplification factor} (WAF), the flash writes per host write~\cite{hu2009fundamental,desnoyers2014analytic}, and take flash time from every application on the SSD~\cite{huang2017flashblox,haas-ssdiq2024}.
Grouping data that die together lets their blocks empty on their own~\cite{van2014necessity}.

NVMe Flexible Data Placement (FDP), part of the standard SSD interface, hands this grouping to the host, but only through a few \emph{handles}, write destinations the SSD keeps open~\cite{edge2025lwnfdp}.
Each write names a handle, and writes to different handles land in different blocks~\cite{nvme-fdp-tp4146,samsung-fdp-whitepaper2024}.
The devices in prior FDP studies~\cite{allison2025fdpcache,son2026iostream} and ours offer only eight at most.
With more kinds of data than handles, some must share one, and FDP pays off only when the data sharing a handle also die together.
The question is therefore who decides which data share each handle, and with what knowledge.

\vspace{2pt}
\noindent\textbf{Motivation.}
Today, a person or a fixed rule decides in advance by drawing a \emph{map} of which data share a handle: \emph{manual assignment}.
Its limits show most clearly on a \emph{shared node}, a server where an operator runs several \emph{tenants}, each an application with its files, on one SSD without changing their code.
There, each existing route covers only part of the decision.
A developer can build a map into one application, as the RocksDB key-value store does.
However, on Linux's common file systems, ext4 and XFS, that map reaches the SSD only through raw device commands that bypass the \emph{page cache}, the kernel's in-memory copy of file data.
An operator can give each tenant a handle of its own, but only in a patched kernel and from a map written by hand~\cite{son2026iostream}.
A flash file system such as F2FS, in turn, applies one fixed rule to all tenants, whatever their data's lifetimes~\cite{qi2026f2fsfdp}.

Section~\ref{sec:motiv} measures three reasons these routes fall short.
First, a \emph{lifetime-aware} map, which groups data by how long they live, needs a costly study.
A \emph{lifetime-blind} map, drawn without one, gains little and can even raise WAF.
Delivering the map from user space also costs the page cache, which made RocksDB's reads 8$\times$ slower.
Second, lifetimes also differ inside a tenant, so one handle per tenant cannot separate them.
Third, even a lifetime-aware map goes stale once the writing tenants change.

\textit{Placement on a shared node is therefore not simply labeling data hot or cold, but an allocation problem.}
The tenants' data fall into more lifetime \emph{classes}, groups of writes that die at about the same time, than there are handles.
Someone must decide which get their own handle.
\textit{That decision needs two facts that exist only at run time: which data die together, and how much GC copying each separation saves on this device, next to these tenants.}
No person sees both: a developer sees one application, an operator only whole tenants.

\vspace{2pt}
\noindent\textbf{Our Insight.}
\textit{The kernel already sees what both facts rest on, because every write and every death passes through it.}
Data die when they are overwritten or \emph{discarded}, when the file system tells the SSD they were deleted.
By charging each death to the file, thread, and tenant that wrote the data, the kernel can learn which data die together.
The share of each class still alive when GC reclaims its block then estimates how much separating that class would save.
\emph{Extended Berkeley Packet Filter (eBPF)}, small verified programs run inside an unmodified kernel, lets us observe and attribute these deaths.

\vspace{2pt}
\noindent\textbf{Our Solution.}
Building on this insight, we present \tbdfixed{} (Figure~\ref{fig:overview}), which performs the allocation for unmodified applications on a \emph{stock} Linux kernel, neither patched nor rebuilt.
It only loads a small module.
To our knowledge, it is the first FDP placement policy that dynamically learns lifetime classes from the deaths an unmodified kernel observes.

Its three components answer those three reasons in turn:
\vspace{-4pt}
\squishlist
\item[\stepB{1}] The \tbdfixedmonitor{} replaces the lifetime study (Section~\ref{sec:design-mon}).
Its eBPF programs label each write by file, thread, and, in a key-value store, the merges its data passed through.
They observe when the write dies, and set its handle on the normal I/O path, keeping the page cache.
\item[\stepB{2}] The \tbdfixedplanner{} turns these measurements into a \emph{layout}, which classes share which handle (Section~\ref{sec:design-planner}).
It gives each tenant a handle first and spends the remaining handles where separation saves the most GC copying.
\item[\stepB{3}] The \tbdfixedloader{} keeps the layout current (Section~\ref{sec:design-loader}).
It installs each layout so that no write ever sees half of one, and plans again when the set of tenants that write changes.
\squishend

Experiments on an FEMU-emulated~\cite{li2018femu} and an enterprise FDP SSD confirm all three answers.
First, labels inferred from kernel events alone did as well as exact labels from a patched RocksDB.
Second, with four databases, \tbdfixed{} removed 45\% of the \emph{excess WAF}, the extra writes GC adds, twice the 22\% that \emph{isolation}, one handle per tenant, remove.
Every writing tenant also ran 19--23\% faster than under isolation.
The same number of handles split differently removed only 37\%, so where handles go matters as much as how many.
Third, on the emulated SSD, after a tenant switch, \tbdfixed{} installed a new layout within about 7\,min with no one redrawing the map.
An hour later WAF stayed a quarter below the old layout's (2.51 vs.\ 3.35).
This costs 8--9\% more CPU per small write, slowing only CPU-bound writers.
\section{FDP and Linux Kernel Mechanisms}
\label{sec:background}

\subsection{NVMe Flexible Data Placement}
\label{sec:fdp}


FDP~\cite{nvme-fdp-tp4146,nvme-basespec-2-1,samsung-fdp-whitepaper2024,zhang2025storage} lets the host choose which writes share flash, while the SSD still runs GC.
The SSD erases flash in groups of erase blocks called \textbf{reclaim units (RUs)}.
It keeps one RU open for each \textbf{reclaim unit handle (RUH)}, which we call a \emph{handle}.
A write names this handle through the \textbf{placement identifier (PID)} in its command.
From this PID, the SSD then maps its placement handle (PHNDL) to one RUH, writing into the corresponding RU (Fig.~\ref{fig:fdp}(a)).
Writes with different handles thus never share an RU.

Separating writes helps because, before it erases an RU, GC must copy every page in it that is still valid, that is, not yet overwritten or deleted.
Thus, how much it copies depends on each page's \emph{lifetime}, the time from its write until it becomes invalid, compared with the \textbf{reclaim age}\,$\hat A$, the time from a page's write until GC reclaims its RU.
Data that die well before $\hat A$ leave nothing to copy, while data that outlive $\hat A$ must be copied.
An RU that mixes the two thus forces GC to copy the long-lived pages.
Separating them lets the RUs of the short-lived data empty on their own~\cite{desnoyers2014analytic,sepbit}.

Separating data by lifetime, however, is constrained by the number of placement handles.
Each handle holds an open RU and consumes device resources such as write-buffer space~\cite{samsung-fdp-blog-2025}. 
Although vendor sources describe up to 128 or 256 handles~\cite{samsung-fdp-blog-2025,samsung-fdp-whitepaper2024}, the devices in published studies and our evaluation expose only 7 or 8~\cite{allison2025fdpcache,son2026iostream}.

This constraint becomes more acute when multiple \emph{tenants}, independently managed workloads, share an SSD.
Each tenant may write data with different lifetimes.
We call writes from the same tenant that die at similar times a \textbf{class}, and define its \textbf{survival} as the fraction of bytes still valid at reclaim age $\hat A$.
Separating classes with different survival reduces GC copying, but the number of classes can exceed the available handles.
The host must therefore identify lifetime classes and allocate scarce handles to the separations that save the most GC copying.

\subsection{Placement in the Linux Write Path}
\label{sec:linuxpath}

Linux provides a mechanism to convey placement decisions to FDP SSDs, but no policy that fits the workload.
Since Linux 6.16, each write bio carries a stream identifier (\texttt{bi\_write\_stream}), which the NVMe driver maps to an FDP PID~\cite{hellwig2024writestreams,busch2025fdpstreams}.
However, the kernel neither measures lifetime classes nor allocates the limited streams among tenants.

Existing approaches require explicit placement decisions, each at a cost (Fig.~\ref{fig:fdp}(b)).
Applications can issue NVMe passthrough commands through \texttt{io\_uring}, but bypass the page cache~\cite{joshi2024passthru}.
F2FS maps hot, warm, and cold data to three streams by one fixed rule, whereas ext4 and mainline XFS provide no such mapping~\cite{qi2026f2fsfdp,joshi2026xfsstreams}.
Operators can assign streams to cgroups, but only through kernel modifications, as in io.stream~\cite{son2026iostream}.

To avoid these changes while preserving the regular I/O path, placement must be enforced at the block layer.
Each write reaches this layer as a bio, where its stream can be set before the NVMe driver constructs the command.
However, kernel flushers may submit buffered writes, whose bios no longer say which file or thread produced them, so a policy must capture that identity earlier.

\subsection{Extended Berkeley Packet Filter (eBPF)}
\label{sec:ebpf}

eBPF lets users add small programs to a running kernel without patching or rebooting it.
It is widely used to trace the kernel, filter network packets, and enforce security policies~\cite{linux-bpf-docs}.
\tbdfixed{} needs it to observe a write's identity where the kernel still knows it, carry that identity to the block layer, and set the stream there (Section~\ref{sec:design-monitor}).

eBPF programs run at \emph{hooks}, fixed points in the kernel, such as a file open or a write entering the block layer~\cite{linux-bpf-docs,linux-libbpf-progtypes-doc}.
They also keep state in key-value \emph{tables} (eBPF's maps) that carry what one hook observed to another.
For example, the label a file gets at open is looked up when its writes reach the block layer.
User programs can read and fill the same tables, so the lifetimes \tbdfixed{} measures go out to user space and its placement decisions come back in (Section~\ref{sec:design-planner}).
This hand-off lets planning, which sorts lifetimes and searches over groupings, run in user space, since the kernel's \emph{verifier} rejects at load time any program that could loop without bound~\cite{linux-bpf-designqa-doc}.
Planning also need not run per write, since a layout changes only when the workload does.

A program at a tracing hook may read kernel data, but never change them~\cite{linux-v7-2-rc4-src}.
As a result, setting the stream needs a \emph{kfunc}, a kernel function exposed to eBPF~\cite{linux-bpf-kfuncs-doc}.
\tbdfixed{} adds one in a 124-line \emph{loadable module}.
This kfunc only writes the stream of the bio it is given, at a hook that runs before the NVMe driver builds the command (Section~\ref{sec:design-monitor}).
None of these pieces requires patching or rebuilding the kernel.

\begin{figure}[!t]
\centering
\includegraphics[width=\columnwidth]{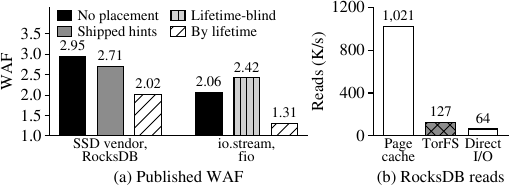}
\caption{
WAF with and without placement by lifetime (a), and RocksDB reads with and without the page cache (b).
(a)\,Published by an SSD vendor~\cite{samsung-fdp-rocksdb-blog} and io.stream~\cite{son2026iostream}.
(b)\,Enterprise SSD, 8\,GiB working set, 2\,GiB block cache.
}
\label{fig:motiv}
\vspace{-8pt}
\end{figure}

\section{Challenges in FDP Placement}
\label{sec:motiv}

Manual FDP placement faces three challenges on shared SSDs, each yielding a requirement: learning lifetimes without losing the page cache (Section~\ref{sec:motiv-manual}), allocating streams to classes finer than a tenant (Section~\ref{sec:motiv-misaligned}), and adapting as the writing tenants change (Section~\ref{sec:motiv-change}).

We study these challenges using RocksDB~\cite{dong2021rocksdb}, PostgreSQL~\cite{postgresql-web}, SQLite~\cite{sqlite-web}, and MongoDB~\cite{mongodb-web}, each running in its own cgroup and directory on a shared ext4 file system backed by an FDP SSD.
Experiments use FEMU~\cite{li2018femu} or an enterprise FDP drive (Section~\ref{sec:eval-setup}). Section~\ref{sec:motiv-misaligned} examines this four-tenant setting, while Sections~\ref{sec:motiv-manual} and~\ref{sec:motiv-change} use one RocksDB instance and a switch between two instances, respectively.
These workloads expose distinct data lifetimes, even within a tenant.
RocksDB writes data into sorted SST files through flushes, and its compactions later rewrite them into deeper \emph{levels} of an LSM tree.
Shallow SST files become invalid within seconds, whereas deeper files survive longer.

%

\subsection{Manual Assignments Are Costly}
\label{sec:motiv-manual}

Assigning streams by hand cuts WAF but brings two problems: a \emph{costly lifetime study} to draw the map, and \emph{slower reads} from bypassing the page cache.

\paragraph{Costly Lifetime Profiling}
The study makes a map \emph{lifetime-aware}, grouping writes by how long they live relative to the reclaim age.
On an SSD vendor's FDP drive, such a \emph{per-level map} reduced RocksDB's WAF about four times as much as its \emph{shipped hints}, lifetime tags chosen by each file's level (Fig.~\ref{fig:motiv}(a))~\cite{samsung-fdp-rocksdb-blog,linux-write-lifetime-hints}.

Achieving these gains, however, requires device-specific reclaim ages and application-specific lifetime labels. 
The vendor compared each level's lifetime with its own drive's reclaim age~\cite{samsung-fdp-rocksdb-blog}, but the reclaim age changes with the device and the load~\cite{desnoyers2014analytic}, by 14$\times$ between our two SSDs.
The required labels also come from application internals: our per-level map relied on a patched RocksDB that tags each SST file with its level (Section~\ref{sec:eval-setup}).

Nor can the study be skipped.
A map drawn without it is \emph{lifetime-blind}.
It groups writes by a fixed rule or a label easier to know than lifetime, and gains little or even raises WAF.
RocksDB's shipped hints group level\,4, which dies near the reclaim age, with level\,5, which outlives it.
On our enterprise SSD, they left WAF at 2.19, against 1.30 for a per-level map (Section~\ref{sec:eval-labels}).
In io.stream's own experiment~\cite{son2026iostream}, 14 cgroups ran fio over seven data streams.
\emph{Hashing} the cgroups onto streams raised WAF from 2.06 with no placement to 2.42, while grouping them by lifetime lowered it to 1.31 (Fig.~\ref{fig:motiv}(a)).
Automatic schemes count rewrites per address as a proxy for lifetime~\cite{yang2017autostream,yong2018vstream,bhimani2022automatic}.
On ext4, however, the file system reuses an address for other files' data, so the count follows the address rather than any file's lifetime.
On RocksDB, placing by this count removed far less excess WAF than lifetime classes did (Section~\ref{sec:eval-labels}).

\paragraph{Page-Cache Bypass}
Reads also slow down because, on ext4 or XFS, an application can apply even a lifetime-aware map only through passthrough, which bypasses the page cache (Section~\ref{sec:linuxpath}).
A read that misses the application's cache then goes to the device, not the page cache.
This detour costs whenever the working set fits in memory but not in that cache.
On the enterprise SSD (8\,GiB working set; 2\,GiB \emph{block cache}, RocksDB's read cache), RocksDB read 8$\times$ faster through the page cache than through TorFS~\cite{torfs-github}, a passthrough FDP plugin.
Direct I/O, which skips only the page cache, was slower still (Fig.~\ref{fig:motiv}(b)).

\finding{1}{
A host policy must learn lifetimes on the node, from what the kernel sees, and place writes on the regular I/O path that keeps the page cache (Requirement \textbf{R1}).
Lifetime-blind maps gain little or even raise WAF, and delivery from user space turns page-cache hits into device reads.
}

\subsection{Classes Within Tenants Outnumber Streams}
\label{sec:motiv-misaligned}

Separating writes by lifetime within tenants creates more classes than available streams, and these classes differ in placement benefit. Yet tenant-level placement, as used by io.stream~\cite{son2026iostream}, cannot distinguish these classes in the first place. A host policy must therefore identify lifetime classes within tenants and allocate scarce streams to the most beneficial splits across the node.


\begin{figure}[!t]
\centering
\includegraphics[width=\columnwidth]{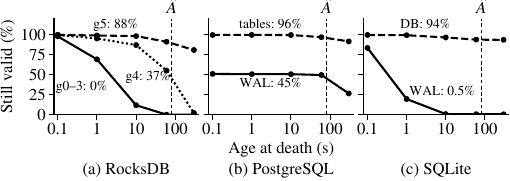}
\caption{
Lifetimes inside the tenants of the four-tenant shared node: share of bytes still valid at each age, measured by \tbdfixed{} (emulated SSD, $\hat A=81.3$\,s).
Labels: survival at $\hat A$; gN: RocksDB generation N (Section~\ref{sec:motiv-misaligned}).
}
\label{fig:lifetime}
\vspace{-6pt}
\end{figure}

\paragraph{Tenant-Level Placement Is Insufficient}
io.stream~\cite{son2026iostream} groups writes by cgroup, treating each tenant as a placement unit. However, a tenant can produce data with widely different lifetimes. Assigning one stream per tenant mixes such data, even when their lifetimes warrant separate placement.

SQLite and PostgreSQL each write a short-lived log and long-lived data. Both first record changes in a write-ahead log (WAL). At each checkpoint, they apply the logged changes to their data and restart the log. 
On our shared node, SQLite's log had a survival of only 0.5\%, compared with 94\% for its database file. 
PostgreSQL's log likewise died much sooner than its tables (Fig.~\ref{fig:lifetime}(b,c)). 
File names distinguish these writes, whereas cgroups do not.

RocksDB requires finer-grained information. Only a file's \emph{generation}, the number of compactions its data has undergone since flush, distinguishes files by lifetime. 
File names alone cannot, since every level produces the same kind of SST file, and a compaction thread writes both shallow and deep files. 
Generations 0--3 die within seconds, generation 4 near the reclaim age, and generation 5 long afterward (Fig.~\ref{fig:lifetime}(a)). 
Because compaction typically moves data about one level deeper, generation tracks level. 
eFDP can infer it from the files read by each compaction (Section~\ref{sec:design-monitor}).

\paragraph{Not All Lifetime Classes Deserve a Stream}
Separating writes within tenants creates more lifetime classes than available streams. Our four tenants form eight classes: a log and a data class each for SQLite and PostgreSQL, three for RocksDB, and one for the idle MongoDB. Yet only seven streams are available.

More importantly, separating these classes yields different benefits. Compared with tenant isolation (one stream per tenant), separating SQLite's log saved 0.25 GC copies per page written (WAF$-1$), separating RocksDB's generations saved 0.16, and separating PostgreSQL's log saved only 0.02 (Section~\ref{sec:eval-alloc}). 
Thus, assigning streams independently within each tenant cannot maximize the benefit of a limited stream budget. The host must prioritize lifetime splits by their expected reduction in GC copying across the node.

\finding{2}{
A host policy must form classes from measured lifetimes, within tenants as well as across them, and allocate scarce streams to the splits that save the most GC copies across the node (Requirement \textbf{R2}). 
Tenant-level placement mixes data with different lifetimes, while finer-grained classes outnumber available streams and differ in placement benefit.
}

\subsection{Lifetime-Aware Maps Become Stale}
\label{sec:motiv-change}

\begin{figure}[t]
\centering
\includegraphics[width=\columnwidth]{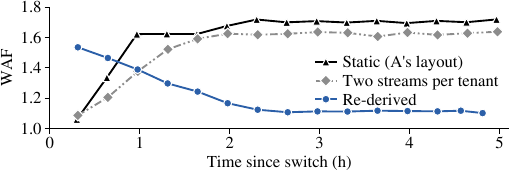}
\caption{
WAF of three maps after a tenant switch, to show what a stale map costs (enterprise SSD, 20-min windows); the load moves from the first RocksDB (A) to the second.
Two streams per tenant: a hand-drawn map giving each RocksDB two of the four streams.
The re-derived run starts from the device state the static run left.
}
\label{fig:stale}
\vspace{-6pt}
\end{figure}

Even a lifetime-aware map can become stale when the active writing tenants change. 
Streams reserved for an idle tenant provide no benefit, while writes from a newly active tenant may mix data with different lifetimes. Thus, \emph{a map can become ineffective even when data lifetimes remain unchanged}.

Consider a workload that switches from one RocksDB instance to another. 
The original map still assigns most streams to the first instance, leaving the second instance's shallow and deep files sharing a single stream. 
On our enterprise SSD, WAF settled at 1.71 after the switch. A static map that reserved two streams for each RocksDB instance offered little improvement (Fig.~\ref{fig:stale}). In contrast, installing a map re-derived offline from the second instance's levels at the switch reduced WAF by 35\%, to 1.11, and improved throughput by 10\%.

\finding{3}{
A host policy must re-derive placement mappings on the node when the active writing tenants change (Requirement \textbf{R3}). 
Static mappings waste streams on idle tenants, while replanning restores the benefits of lifetime-aware placement.
}


\subsection{Requirements and Existing Systems}

Our observations establish three requirements for effective FDP placement: transparent lifetime profiling on the regular I/O path (R1), benefit-aware stream allocation across tenants (R2), and replanning as active writing tenants change (R3).

No existing FDP system satisfies all three requirements (Table~\ref{tab:sota}).
Existing approaches neither learn lifetimes on an unmodified software stack nor independently allocate streams across tenants. 
\tbdfixed{} addresses R1--R3 without modifying the kernel source (Section~\ref{sec:design}). 
It automatically replans when the active writing tenants change, while a new device or workload change requires an operator-triggered replan.

\begin{table}[!h]
\caption{
Existing FDP systems checked against an unmodified stack and R1--R3.
\tyes{}: yes; \tpart{}: only via rules, labels or assignments a person supplies; \tsome{}: partly, in some modes or through a lifetime proxy; \tno{}: no.
Stock kernel: no patch or rebuild (\tbdfixed{} loads a 124-line kfunc module).
}
\label{tab:sota}
\centering
\scriptsize
\setlength{\tabcolsep}{4pt}
\begin{tabularx}{\columnwidth}{@{} l *{7}{c} @{}}
\toprule
 & \multicolumn{3}{c}{Unmodified} & R1 & \multicolumn{2}{c}{R2} & R3 \\
\cmidrule(lr){2-4}\cmidrule(lr){5-5}\cmidrule(lr){6-7}\cmidrule(l){8-8}
System & \makecell{App\\as is} & \makecell{Stock\\kernel} & \makecell{Page\\cache} & Lifetimes & \makecell{Divides\\tenant} & Allocates & Adapts \\
\midrule
TorFS~\cite{torfs-github} & \tno & \tyes & \tno & \tpart & \tyes & \tno & \tno \\
CacheLib FDP~\cite{allison2025fdpcache} & \tno & \tyes & \tno & \tpart & \tyes & \tpart & \tno \\
SlimIO~\cite{lee2025slimio} & \tno & \tyes & \tno & \tpart & \tyes & \tno & \tno \\
RocksDB hints~\cite{samsung-fdp-rocksdb-blog} & \tsome & \tsome & \tyes & \tpart & \tyes & \tno & \tno \\
XFS streams~\cite{joshi2026xfsstreams} & \tno & \tno & \tyes & \tpart & \tyes & \tno & \tno \\
FDPFS~\cite{chen2024fdpfs} & \tsome & \tyes & \tsome & \tpart & \tyes & \tpart & \tno \\
io.stream~\cite{son2026iostream} & \tyes & \tno & \tyes & \tpart & \tno & \tpart & \tno \\
F2FS~\cite{qi2026f2fsfdp} & \tyes & \tyes & \tyes & \tsome & \tyes & \tno & \tno \\
\midrule
\textbf{\tbdfixed{}} & \tyes & \tyes & \tyes & \tyes & \tyes & \tyes & \tyes \\
\bottomrule
\end{tabularx}
\end{table}

\section{Design}
\label{sec:design}

We first formulate the placement problem and outline eFDP's feedback loop. We then describe its three components: the \tbdfixedmonitor{}, which profiles write lifetimes (R1); the \tbdfixedplanner{}, which allocates limited placement streams (R2); and the \tbdfixedloader{}, which installs and updates placement mappings as workloads change (R3).

\subsection{Problem Formulation and Overview}
\label{sec:design-problem}

\begin{figure*}[t]
\centering
\includegraphics[width=0.94\textwidth, keepaspectratio]{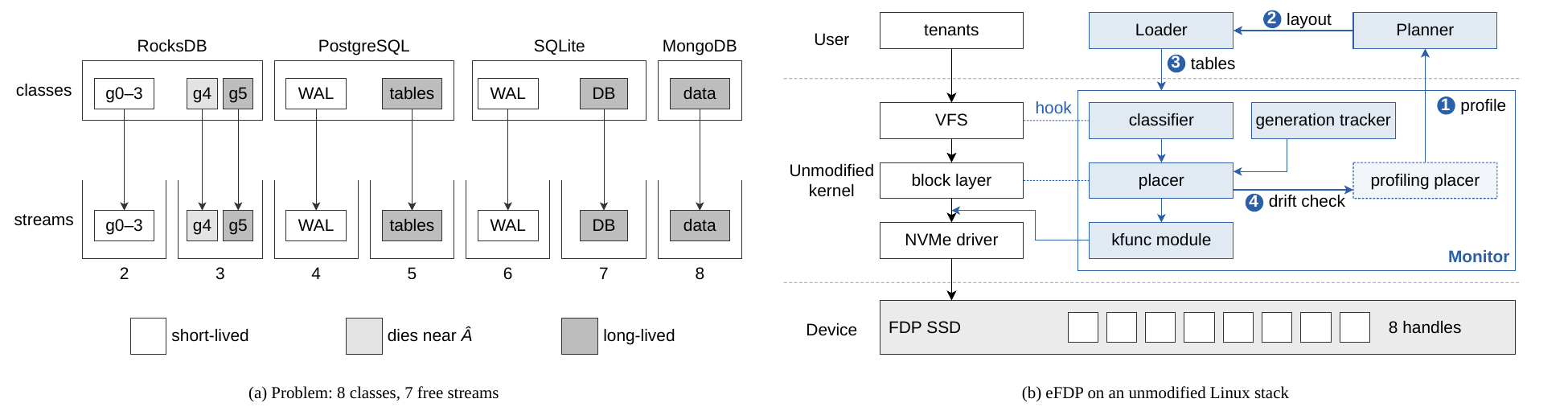}
\caption{
Overview of \tbdfixed{}: the problem it solves (a) and its loop (b).
(a)\,Four tenants' writes form eight classes for seven free streams, so two classes, here g4 and g5, share one.
(b)\,The loop (shaded) on an unmodified Linux stack: profile (\stepA{1}), layout (\stepA{2}), tables (\stepA{3}), drift check (\stepA{4}).
}
\label{fig:overview}
\vspace{-12pt}
\end{figure*}

Fig.~\ref{fig:overview}(a) draws the problem on our shared node, one part per requirement.
Each tenant, a cgroup plus a directory, writes data that fall into classes.
The file and thread that tell its writes apart, however, do not give those classes or their survival.
\tbdfixed{} must thus combine them with the overwrites and discards it observes (R1).
The four tenants form eight classes for seven free streams, so which two share one is an allocation (R2).
Here, RocksDB's generations 4 and 5 (g4, g5) share, because splitting them is worth less than any split that got a stream (Section~\ref{sec:eval-alloc}), although it pays on RocksDB alone (Section~\ref{sec:eval-labels}).
The layout reflects the tenants observed while profiling, so a change of active tenants can leave it unfit and triggers a re-plan (R3).

\problem{
$N$ tenants share an FDP SSD with $B$ free streams.
Tenant\,$i$'s writes fall into $k_i$ classes.
Class\,$c$ has $b_c$ bytes, a fraction $s_c$ still valid at the reclaim age $\hat A$.
Classes and $s_c$ must be learned, and $\sum_i k_i$ may exceed $B$.
Choose which classes share a stream, never across tenants unless the operator waives it, to minimize the live bytes GC copies.
Change no application, file system, or kernel source.
Choose again when the writing tenants change.
}

\tbdfixed{} solves this problem with a feedback loop (Fig.~\ref{fig:overview}(b)).
The \emph{placer}, an eBPF program in the block layer, sets every write's stream.
During a \emph{profiling window}, the \tbdfixedmonitor{} also records when writes die.
From this \emph{profile} of bytes and ages at death (\stepA{1}), the \tbdfixedplanner{} derives a \emph{layout}: file and thread rules plus a class-to-stream table (\stepA{2}).
The \tbdfixedloader{} installs it in seconds (\stepA{3}), and a check once per reclaim age restarts the loop when the writing tenants change (\stepA{4}).

\subsection{\tbdfixedmonitor{}}
\label{sec:design-mon}

The \tbdfixedmonitor{} must tell the \tbdfixedplanner{} which kind of write each write is and when it dies (R1).
However, at the block layer, where streams are set, a write has lost most of its identity, and its lifetime shows only at death, long after.
The \tbdfixedmonitor{} therefore keeps each write's identity on three \emph{signals} the kernel sees, its file pattern, thread name, and generation, and links deaths to writes.
It calls a write's value on a signal its \emph{label}, e.g., \texttt{*.sst}.
We follow one write, W1: RocksDB's flush thread writing the new SST file \texttt{000123.sst}.

\subsubsection{Reconstructing Write Identity}
\label{sec:design-monitor}

The block layer usually keeps only a write's tenant, which the placer needs to keep tenants apart.
Hence, the other labels must be captured where they are still known and joined at the bio.
The bio is usually submitted by a writeback thread, so it carries neither the path nor the writing thread (Section~\ref{sec:linuxpath}).
As the Linux kernel charges memory per page but attributes writeback per file~\cite{linux-cgroupv2-doc}, it assigns each file's \emph{inode}, the kernel's record of the file, to the cgroup that modifies most of its cached pages.
Every bio that writes the file back carries that cgroup.
Since tenants write only their own files, the cgroup names the tenant.

The file label is therefore taken at open, the last point that knows the path.
It is then stored, so that no write resolves its path again.
When a regular file is opened for writing, an eBPF \emph{classifier} matches its path against the layout's file patterns, e.g., \texttt{*.sst} for \texttt{000123.sst}, and records the result for the file.
The \tbdfixedplanner{} makes it a rule only if a pattern's \emph{purity}, the share of its profiled bytes that fall in its most common class, is at least 80\% (Table~\ref{tab:params}).
During profiling, a second program records each path at open, so the \tbdfixedplanner{} never scans directories.
Both programs use under 2\% of the verifier's limit.

The \tbdfixedmonitor{} infers an SST file's level~\cite{oneil1996lsm,dong2021rocksdb}, which neither the path nor the writing thread reveals, from what the writer read.
A \emph{generation tracker} records which SST files each thread opens for reading, and a new file gets one more than the lowest generation its thread read.
A flush reads none, so its file gets generation\,0.
Because a compaction usually writes one level below its shallowest input, generation follows level.

At the block layer, the placer uses small, fixed tables to join the tenant from the bio's cgroup, the file's labels, and the submitting thread's name.
A buffered write finds the file's labels through the page being written.
Direct I/O has no such page, so a small program at submission records them for the request instead.
Writes that kernel writeback threads submit get no thread label.
Bounding the tables, at most 32 file-pattern rules, 8 thread rules, 8 classes per signal, and 16 tenants, lets the verifier accept the loops over them, and every layout in Section~\ref{sec:eval} fits.

With these labels, W1 reaches its handle on the normal I/O path.
As RocksDB writes SST files with direct I/O or syncs them from the writing thread, W1's bio comes from the flush thread with labels \texttt{*.sst}, \texttt{rocksdb:high} and generation\,0.
The class-to-stream table then maps it to stream\,2.
Because the placer's hook runs before the NVMe command exists (Section~\ref{sec:ebpf}), it sets the stream through our kfunc, which skips any bio that already has one.
As a result, a stream set by another layer is never overridden.
The unmodified NVMe driver then puts stream\,2's PID into the command.

\subsubsection{Attributing Deaths to Writes}
\label{sec:design-profile}

A page dies long after its write, often in another thread, but at the same disk address.
The \tbdfixedmonitor{} uses this to link the two.
A lifetime is thus the time from a write the kernel sees to the overwrite or discard of its address.
During a profiling window, a \emph{profiling placer} replaces the placer and stores, for the disk range each write starts in, a 64-byte \emph{owner record} of its file, thread, tenant, class, and time.
A later write or discard over that range means the old bytes died, and they are charged to the recorded owner.
When compaction deletes W1's file, a kernel thread issues the discard, yet the death is charged to \texttt{000123.sst} and RocksDB's flush thread.

Two guards keep these charges nearly correct and their memory bounded.
First, each record notes its file's creation time, so a reused inode never inherits an old file's deaths.
Second, each \emph{owner range}, 64\,KiB or capacity$/2^{22}$ if larger (Table~\ref{tab:params}), keeps only its latest owner, so the records never exceed 256\,MiB.
A write that overlaps a record charges its owner's whole extent.
A newer owner in the same range displaces an older one, whose later deaths go uncounted.

What the \tbdfixedmonitor{} hands the \tbdfixedplanner{} is small: one histogram of age at death per label, whose six bins, with edges at 0.1, 1, 10, 60 and 300\,s, bracket the emulated SSD's reclaim age.
The last bin also holds bytes never seen dying, which count as alive.
When a window ends, a user-space agent reads the histograms through the \texttt{bpf()} system call.
It then hands them, with the paths captured at open, to the \tbdfixedplanner{}.

\subsection{\tbdfixedplanner{}}
\label{sec:design-planner}

To meet R2, the \tbdfixedplanner{} must decide which splits of a tenant's data save the most GC copying and which get the few streams.
Each split is a \emph{boundary}, dividing one tenant's labels on one signal into two classes.
The profile usually shows more boundaries than the SSD has streams, and the \tbdfixedplanner{} gets no help from the device, which never reports what a split would save.
Since sorting, floating-point, and open-ended search are hard in eBPF and unneeded per write, it runs in user space once per layout.
This in turn leaves about ten table lookups per write in the kernel.

\subsubsection{Finding and Valuing Lifetime Boundaries}

A boundary helps only where lifetimes cross the reclaim age, because GC copies what is still valid then (Section~\ref{sec:fdp}).
The \tbdfixedplanner{} therefore scores each set of labels $X$ by its \emph{survival rate},
\begin{equation}
    s_X=1-d_X/b_X,
    \label{eq:survival}
\end{equation}
the share of the $b_X$ bytes written in the window still valid at the \emph{scoring age}, $\hat A$ by default (Section~\ref{sec:eval-sens} varies it).
The bytes dead by then, $d_X$, come from the age bins.\footnote{Spread log-uniformly within each bin. Assuming a steady write rate, a death at age $a$ in a window of length $W$ is weighted by $W/(W-a)$, because only bytes written in the window's first $W-a$ could die at that age within it.}
It estimates $\hat A$ as capacity over the window's media write rate, from the media-written counter of the device's \emph{FDP statistics log}.
Unlike median age at death, survival separates a class that dies just before $\hat A$ from one that outlives it (Section~\ref{sec:eval-labels}).
Candidate boundaries lie only between labels that are neighbours in survival (for generations, in number).
Each side requires at least 5\% of the tenant's bytes, the \emph{minimum side size}, as smaller sides save little (Section~\ref{sec:eval-sens} tests 1--20\%).

From survival, the \tbdfixedplanner{} estimates a boundary's \emph{worth}, the GC copies it would save,
\begin{equation}
    w=b_L\,s_L\,(s_L-s_S),
    \label{eq:worth}
\end{equation}
where $L$ is the longer-lived side and $S$ the shorter-lived one.
Intuitively, if both sides share RUs, the shorter-lived side's deaths make units eligible for GC~\cite{rosenblum1992design,desnoyers2014analytic}.
The longer-lived side's pages in them are still valid, which GC later copies.
$b_L s_L$ bounds those copies, and the survival gap thus approximates how many units GC picks.
The model assumes \emph{greedy GC}, which reclaims the unit with the fewest valid pages, and a steady load.
Section~\ref{sec:eval-alloc} checks its ranking.

\begin{algorithm}[t]
\caption{Tenant-first stream allocation.}
\label{alg:planner}
\footnotesize
\renewcommand{\algorithmicrequire}{\textbf{Input:}}
\renewcommand{\algorithmicensure}{\textbf{Output:}}
\begin{algorithmic}[1]
\algnotext{EndLoop}
\Require each tenant's labels per signal (file, thread, generation), sorted by survival $s$ at $\hat A$ (generations by number); stream budget $B$
\Ensure file and thread rules; a class-to-stream table per tenant
\State Give each tenant its own stream; if tenants $>B$, the $B-1$ largest by bytes get one each and the rest share one
\State $\mathit{free}\gets B-{}$streams used in line~1
\Loop
\State Per tenant and signal, find the allowed boundary ($\geq$5\% of the tenant's bytes per side) of highest worth $w$; $\kappa\gets{}$streams it adds
\State Discard boundaries with $\kappa>\mathit{free}$ or that exceed a rule limit
\State \textbf{if} none remain \textbf{then break}
\State Commit the boundary with the highest $w/\kappa$; $\mathit{free}\gets\mathit{free}-\kappa$
\EndLoop
\State Number each tenant's classes by survival; class~0 (longest-lived) uses the tenant's own stream and takes writes that match no rule
\State Rules are shared across tenants: if two map one file pattern or thread name to different classes, keep the mapping of the tenant that wrote more bytes under it
\State Output the rules and class-to-stream table; reject the layout if a class can match no write
\end{algorithmic}
\end{algorithm}

\subsubsection{Allocating the Stream Budget under Tenant Isolation}

Algorithm~\ref{alg:planner} first gives each tenant its own stream, \emph{tenant-first allocation}, then spends the remaining streams by worth per added stream, $w/\kappa$, within the rule limits.
Worth alone cannot decide which boundaries get streams, since a write's class combines its labels on every signal.
If RocksDB already has two thread classes, a boundary between its generations makes $2\times2=4$ classes and costs $\kappa=2$ streams.

Tenant-first is a guarantee the operator chooses.
With at most $B$ tenants (beyond that, the overflow tenants share one stream without it; Algorithm~\ref{alg:planner}, line\,1), an installed layout never sends two tenants' writes to one stream, so no RU the host fills mixes tenants.
The guarantee stops at the host.
Writes share the first handle during a swap (Section~\ref{sec:design-loader}), and the enterprise SSD's handles are \emph{initially isolated}, so its GC may combine valid data from different handles.
Where streams are scarce, the guarantee costs WAF, e.g., a stream kept for an idle tenant (Section~\ref{sec:eval-alloc}).
An operator who puts WAF first can instead cluster the same labels with k-means, ignoring tenants, after PCStream's clustering~\cite{kim2019fully}, though not its program-context extraction.

\begin{table}[t]
\caption{
\tbdfixed{}'s parameters, their defaults, and how each is set.
$B$ and $\hat A$ are measured; the layout does not depend on the minimum side size, purity, or scoring age over wide ranges (Section~\ref{sec:eval-sens}).
}
\label{tab:params}
\centering
\scriptsize
\setlength{\tabcolsep}{3pt}
\renewcommand{\arraystretch}{1.05}
\begin{tabularx}{\columnwidth}{@{}>{\raggedright\arraybackslash}p{1.45cm}>{\raggedright\arraybackslash}p{1.75cm}>{\raggedright\arraybackslash}X@{}}
\toprule
Parameter & Default & How it is set \\
\midrule
Streams $B$ & 7 (emulated) & Driver's streams, minus one for unplaced writes \\
Reclaim age $\hat A$ & 81--96\,s; ent.\ $\approx$23\,min & Capacity $\div$ flash write rate (host writes + GC copies) during profiling \\
Window $W$ & $4\hat A$ & Longer windows change no stream assignment \\
Drift check & every $\hat A$ & A tenant is active if it wrote $\geq$5\% of all tenants' bytes in the last $\hat A$; trigger delay $\leq\hat A$ \\
Min.\ side size & 5\% & Each side holds $\geq$5\% of the tenant's bytes \\
File-pattern purity & 80\% & A pattern becomes a rule only if $\geq$80\% of its bytes fall in one class \\
Owner range & 64\,KiB & $\max(64\,\mathrm{KiB}, \mathrm{capacity}/2^{22})$, a power of two; one 64-byte record each, $\leq$256\,MiB \\
\bottomrule
\end{tabularx}
\vspace{-6pt}
\end{table}

\subsection{\tbdfixedloader{}}
\label{sec:design-loader}

The \tbdfixedloader{} must install each layout safely and keep it matched to the writing tenants.
Safety means no write sees a half-installed layout, no layout names more streams than the device has, and scarcity never moves one tenant's class onto another's handle.
Since a layout fits only the tenants it was profiled on (Section~\ref{sec:motiv-change}), R3 asks for a new one when they change.

\subsubsection{Installing a Layout Safely}

The \tbdfixedloader{} installs a layout so that all three safety conditions hold.
It rejects any layout that uses more streams than the driver reports or, with at most $B$ tenants, leaves a tenant without its own stream.
It then attaches the programs and fills their tables, writing the on/off flag last, so no write sees half a layout.
At runtime, a class without a stream falls back to its tenant's own stream, so no class lands on another tenant's handle.
The classifier misses files opened before loading, so one scan of the tenants' directories labels them.

A re-plan replaces a layout by the same procedure rather than editing it in place, so no write is ever placed by a mix of old and new tables.
The \tbdfixedloader{} unloads the running programs with their tables, then installs the new layout as above.
Writes in between carry no stream and go to the first handle.
For these seconds, only the tenants share it, as they would without \tbdfixed{}, and nothing already written moves.
The scan then labels the files that stayed open.
Generations live in the tables too, so a swap discards them.
The swap keeps the table of generations, so files written before it keep theirs.

\subsubsection{Re-Planning When Tenants Change}

\tbdfixed{} re-plans only when the set of active tenants changes (R3).
On every $\hat A$, a \emph{drift check} reads the per-tenant byte counts that the placer already keeps.
It calls a tenant \emph{active} if it wrote at least 5\% of all tenants' bytes in the last $\hat A$.
In our switch runs, the new writer's share rose from 0 to about half, well past this threshold (Section~\ref{sec:eval-adapt}).
A new tenant that writes less than 5\% is not caught.
A load change among the same tenants, or a drift in $\hat A$, triggers no re-plan; an operator can start one.

A re-plan costs one profiling window $W=4\hat A$, since longer windows changed no stream assignment in our profiles (Section~\ref{sec:eval-sens}).
The \tbdfixedloader{} swaps in the profiling placer for $W$.
The \tbdfixedplanner{} then computes a new layout under the same budget, which the \tbdfixedloader{} installs with the normal placer.
Writes go unplaced only during the swaps themselves.
Because a re-plan needs a new set of active tenants and the next check follows the install, re-plans start at most once every $W+\hat A$, even when a tenant hovers at the threshold.

\section{Evaluation}
\label{sec:eval}

\subsection{Experimental Setup}
\label{sec:eval-setup}

\textbf{Implementation.}
\tbdfixed{} runs on an unmodified Linux kernel.
Its eBPF programs attach at kernel function entries and tracepoints, and set streams through our 124-line kfunc module.
They take about 1,500 lines of C, and its user-space agent, \tbdfixedplanner{}, and \tbdfixedloader{} about 4,000.
The agent and Loader load the programs through libbpf.
We evaluate it on Linux 7.2-rc4 with a FEMU-emulated FDP SSD and on Linux 7.0 with a real enterprise FDP SSD (Table~\ref{tab:setup}).

We evaluate four workloads (Table~\ref{tab:setup}):
\squishlist
\item \textbf{Single RocksDB}: one RocksDB instance.
\item \textbf{Shared node}: the four databases from Section~\ref{sec:motiv}, sharing 2--7 streams.
\item \textbf{Tenant switch}: the write load shifts from one RocksDB instance to another.
\item \textbf{Mixed node}: the shared node with an additional kernel-build loop and a 4\,KiB direct random reader.
\squishend
All datasets are initially loaded without placement. Placement begins with subsequent overwrites. We report fill as used space relative to ext4's usable capacity, excluding its 5\% root reserve.

\begin{table}[tp]
\caption{
Testbeds and workloads of the evaluation.
}
\label{tab:setup}
\centering
\scriptsize
\setlength{\tabcolsep}{3pt}
\renewcommand{\arraystretch}{1.1}
\begin{tabularx}{\columnwidth}
{@{}>{\raggedright\arraybackslash}p{1.0cm}>{\raggedright\arraybackslash}X>{\raggedright\arraybackslash}p{3.25cm}@{}}
\toprule
 & Emulated & Enterprise \\
\midrule
Device & FEMU 9.0 (QEMU 9.0.1)~\cite{li2018femu} with FDP; 32\,GiB namespace; 8 handles, seven free; RU 128\,MiB (256); 8$\times$8 LUNs, 4\,KiB pages, read/prog./erase 40/200\,\textmu s/2\,ms & NVMe FDP SSD, 3.84\,TB; 3.2\,TiB namespace; 8 handles, seven free; RU 12\,GiB (270) \\
Host & 8 vCPUs, 15\,GiB; KVM, 16-core x86; Linux 7.2-rc4, unmodified & 8-core x86, 2.9\,GHz, 24\,GiB; Linux 7.0 \\
FS, WAF & ext4, online discard; flash/host pages, last 900 of 1,200\,s & ext4 or F2FS; media/host bytes \\
Software & \multicolumn{2}{>{\raggedright\arraybackslash}p{7.3cm}@{}}{RocksDB 9.10 (db\_bench), PostgreSQL 16 (pgbench), MongoDB 8.0, SQLite 3.45 (WAL mode), fio 3.36 (emulated only)} \\
\bottomrule
\end{tabularx}

\vspace{4pt}
\begin{tabularx}{\columnwidth}{@{}>{\raggedright\arraybackslash}p{1.0cm}>{\raggedright\arraybackslash}X>{\raggedright\arraybackslash}p{1.45cm}>{\raggedright\arraybackslash}p{0.45cm}@{}}
\toprule
Workload & Tenants (runs 1,200\,s unless noted) & Fill (\%) & $B$ \\
\midrule
One RocksDB & 22\,M 1\,KiB keys, overwrite, one thread; ent.\ 2.5\,G, hours & 74; ent.\ 77, F2FS 76 & 3--5 \\
Shared node & RocksDB 5.6\,M keys, PostgreSQL scale 641, SQLite 2.4\,M rows overwrite; MongoDB 6.4\,M docs idles & 72 & 2--7 \\
Switch & two RocksDB taking turns, 9.8\,M keys each, 20 + 30 or 60\,min; ent.\ 1.25\,G, hours & 65; ent.\ 85 & 4 \\
Mixed node & shared node at 0.92$\times$ size, kernel build, 4\,KiB reader & 72 & 7 \\
\bottomrule
\end{tabularx}
\vspace{-12pt}
\end{table}

We compare \tbdfixed{} with no placement (all writes on the first handle) and with each kind of map in \S\ref{sec:motiv}:
\squishlist
\item \textbf{Isolation}: one stream per tenant, io.stream's map.
\item \textbf{Developer maps}: RocksDB's shipped hints and an exact per-level map from a RocksDB patched to tag each SST file by level, as in SlimIO.
\item \textbf{Automatic maps}: AutoStream's rule~\cite{yang2017autostream}, re-implemented at the same block-layer hook, which streams each write by its 1\,MiB range's decaying write count (halved every 10\,s);
and k-means as in PCStream~\cite{kim2019fully}, which clusters \tbdfixed{}'s labels by lifetime alone.
\item \textbf{Systems with their own I/O path}:  TorFS~\cite{torfs-github} and F2FS's temperatures~\cite{qi2026f2fsfdp}.
\squishend
As io.stream demands a patched kernel and SlimIO a patched database~\cite{son2026iostream,lee2025slimio}, we run their maps in \tbdfixed{}'s placer.

We report WAF and the share of the excess WAF that a layout removes, $(W_0-W)/(W_0-1)$, where $W_0$ is the WAF without placement.
Ext4's journal writes stay on the first handle under every layout, so any placement gains about 0.12 WAF by leaving it.
We therefore credit a boundary only against another placed layout.
On the emulated SSD, runs were paired and their order randomised.
Identical configurations differed by 0.025 WAF, so we call a difference real only when its 95\% interval lies beyond $\pm0.025$, and negligible under 0.1 WAF.

\subsection{Can Inferred Labels Replace an Application's Own?}
\label{sec:eval-labels}

\begin{figure}[t]
\centering
\includegraphics[width=\columnwidth]{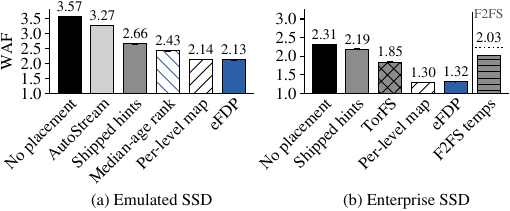}
\caption{
WAF of one RocksDB under each placement, testing inferred labels against RocksDB's exact levels (a) and the Planner's three streams against a five-stream per-level map (b).
(a)\,Emulated SSD, 128\,MiB units, 74\% fill.
(b)\,Enterprise SSD, 77\% fill (F2FS 76\%); the Planner is given RocksDB's levels, testing only its grouping.
Right of the rule: F2FS; dotted: F2FS without placement.
Whiskers span the runs.
}
\label{fig:labels}
\vspace{-12pt}
\end{figure}

Labels inferred from kernel events alone did as well as RocksDB's exact levels.
Given the vendor's grouping, $\{0\text{--}3\}\{4\}\{5\}$ (levels 0--3 together, 4 and 5 apart), \tbdfixed{} sent nearly the same bytes to each stream as the exact per-level map, and matched its WAF, 2.13 against 2.14 (Fig.~\ref{fig:labels}(a)).

The \tbdfixedplanner{} also places the boundaries well, whereas rules that rank labels by another statistic can misplace one.
The \tbdfixedplanner{} scores labels at the reclaim age, here about 96\,s, where its three classes die before, near, and after reclaim (\S\ref{sec:motiv-misaligned}).
Ranking by median age at death instead merges generations\,4 and\,5, so generation\,5's live pages sit in units that generation\,4's deaths empty.
As a result, WAF rose by 0.31 in paired runs.
RocksDB's shipped hints make the same merge, at 0.53 above \tbdfixed{}.
AutoStream's address counts instead mix levels.
On four streams each, it removed 12\% of the excess WAF to \tbdfixed{}'s 56\%.

The same grouping carries over to the enterprise SSD, whose reclaim age, about 23\,min, is 14$\times$ the emulator's.
Given RocksDB's true levels and their ages at death profiled on the device, so that only the grouping was tested, the \tbdfixedplanner{} again chose the vendor's three classes, at any scoring age from 0.5 to 2\,$\hat A$.
On three streams, it matched a five-stream per-level map, 1.32 against 1.30 (Fig.~\ref{fig:labels}(b)).
Fixed rules removed far less of each file system's excess WAF: 9\% for the shipped hints on ext4 and 16\% for F2FS's temperatures on F2FS,\footnote{We test F2FS only on the enterprise SSD. On the emulated SSD, F2FS with its temperature placement and segment-sized discard had over twice the WAF of ext4 without placement.} against 76\% for the \tbdfixedplanner{} on ext4.
This is because F2FS folds RocksDB's hints into three temperatures and thus puts level\,3 with all deeper levels.

\subsection{Is Isolating Tenants Enough?}
\label{sec:eval-shared}

Isolating tenants is not enough, as boundaries inside tenants double the excess WAF that isolation removes.
An operator who does not study the workloads gives each tenant one stream, as io.stream does.
That map captures the whole gain only if lifetimes differ only between tenants, and it leaves three of our seven streams idle.
We compare isolation with no placement and with two seven-stream \tbdfixed{} layouts from one profiling window.
Both keep PostgreSQL's and SQLite's logs apart from their data and differ only in RocksDB's boundary.
\emph{\tbdfixed{}-G}, which Algorithm~\ref{alg:planner} chooses, splits generations\,0--3 from deeper ones.
Meanwhile, \emph{\tbdfixed{}-T} spends that stream on RocksDB's flush thread.

\begin{figure}[t]
\centering
\includegraphics[width=0.9\columnwidth]{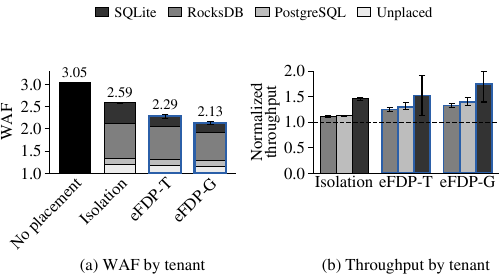}
\caption{
Per-tenant WAF (a) and throughput (b) on the four-tenant shared node, comparing isolation with boundaries inside tenants.
Emulated SSD, 128\,MiB units.
(a)\,GC copies per page written, by tenant, so each bar's top is its WAF; without placement, copies cannot be attributed to a tenant.
(b)\,Throughput normalised to no placement (dotted line).
\tbdfixed{}-T and \tbdfixed{}-G divide RocksDB by flush thread and by generation.
Whiskers span min--max.
}
\label{fig:shared}
\vspace{-6pt}
\end{figure}

Isolation removed 22\% of the excess WAF, and \tbdfixed{}-G twice that, 45\% (Fig.~\ref{fig:shared}(a)).
Since GC copies take flash time from all tenants, fewer copies also let RocksDB, PostgreSQL, and SQLite run 19--23\% faster than under isolation (Fig.~\ref{fig:shared}(b)).

Most of the gain over isolation comes from a boundary no one had to name: SQLite's log.
From isolation to \tbdfixed{}-G, GC copies of SQLite's data per page written halved, from 0.463 to 0.217, more than half of the node's drop.
The log dies long before the reclaim age while the database outlives it (Section~\ref{sec:motiv-misaligned})~\cite{sqlite-wal}.
As a result, when they share a stream, every unit that the log's deaths empty still holds database pages to copy.
The profile found this log unaided, whereas a per-file rule~\cite{fstream,valet} needs an operator who knows it.

The stream count alone does not explain the gain, because \tbdfixed{}-T used as many streams as \tbdfixed{}-G, yet removed only 37\%.
RocksDB's flush thread writes only level-0 files, while its compaction threads write short-lived shallow and long-lived deep levels alike.
Splitting by thread left GC copies of RocksDB's data near isolation's, 0.74 against 0.79 per page written.
Only the generation split separates shallow from deep levels, lowering WAF by a further 0.155, three-quarters on RocksDB's streams.

In our mixed node, non-database tenants did not displace the main boundary, as SQLite's log again got its own stream.
With six tenants and seven streams, the \tbdfixedplanner{} gave each tenant its own stream, leaving none for RocksDB's generations.
With databases scaled to keep the same fill, this layout removed 35\% of the excess WAF, from 2.66 to 2.09.
RocksDB and PostgreSQL ran 16\% and 26\% faster and the reader's throughput rose 22\%, while the build ran within 5\% of its rate without placement.
SQLite's change was within noise.

\subsection{Who Gets a Stream When Streams Run Out?}
\label{sec:eval-alloc}

\begin{figure}[t]
\centering
\includegraphics[width=\columnwidth]{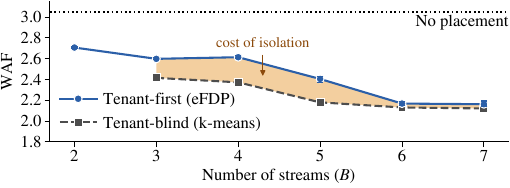}
\caption{
WAF of tenant-first and tenant-blind allocation by stream budget, to show what keeping each tenant on its own streams costs (shaded).
At our SSDs' seven free streams, the cost is under 0.1 WAF.
Shared node, emulated SSD, one profile; mean of 2--3 runs, min--max whiskers.
}
\label{fig:budget}
\vspace{-6pt}
\end{figure}

When streams run out, \tbdfixed{}'s default, tenant-first allocation, costs almost nothing at seven streams.
Fig.~\ref{fig:budget} sweeps the budget $B$ on the shared node against \emph{tenant-blind k-means}.
At seven streams, the gap was negligible, 2.12 against 2.16, even with the tenant guarantee and generations\,4 and\,5 kept together (Fig.~\ref{fig:overview}(a)), while k-means gave generation\,4 its own stream.

Under tenant-first allocation, each stream beyond the four tenant streams went where it saved the most, as worth predicted.
The fifth went to SQLite's log and the sixth to RocksDB's generations, which together lowered WAF by about 0.45, while the seventh, PostgreSQL's log, changed it by less than 0.1.
The three boundaries were worth 9.3\,GiB, 6.8, and 4.0.
From isolation to \tbdfixed{}-G, their tenants' GC copies per page written fell by 0.25, 0.16, and a negligible 0.02.
Splitting generation\,4 from\,5, worth about 2.6\,GiB, ranked last and got no stream.
Overall, WAF fell from 2.71 at two streams to 2.16 at seven.

Tenant-first allocation trails a tenant-blind rule only when streams are scarcer, and then by about 0.2.
With three to five streams, k-means was 0.18--0.24 below tenant-first.
This gap reflects both the tenant guarantee and the different grouping that k-means draws.
At four and five streams, the guarantee spent one stream on the idle MongoDB.
Scarcity, however, comes quickly.
With two or three classes per tenant, three writing tenants already fill seven streams.

Where lifetimes separate cleanly, the two rules agree.
On one RocksDB, k-means derives \tbdfixed{}'s layout byte for byte.

\subsection{When Does the Layout Change, and What Does It Recover?}
\label{sec:eval-adapt}

\begin{figure}[t]
\centering
\includegraphics[width=\columnwidth]{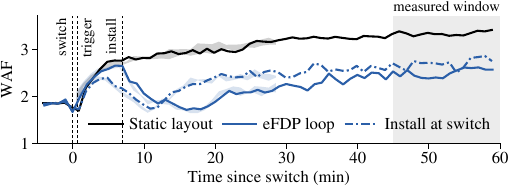}
\caption{
WAF per minute across a tenant switch, comparing \tbdfixed{}'s loop, a static layout, and the loop's grouping installed at the switch.
Emulated SSD, four streams; the first RocksDB writes for 20\,min, then the second.
Light bands: min--max of three earlier 30-min runs; shaded: the measured window.
}
\label{fig:adapt}
\vspace{-6pt}
\end{figure}

\tbdfixed{} re-plans only when the writing tenants change, within about 7\,min, and then beats a static layout.
On the emulated SSD, the write load moved from the first RocksDB to the second (Fig.~\ref{fig:adapt}).
Every run started from the first one's layout, which puts all of the second's data on one stream.
We compare \tbdfixed{}'s loop with a \emph{static layout} that keeps the first layout, as a map drawn once would.
We then compare it with \emph{install-at-switch}, which installs the loop's grouping at the switch to show what waiting for a profile costs.
Each ran three times for 30\,min after the switch and once for 60\,min.

The drift check, run every $\hat A=89$\,s, caught the switch at its first check, 42\,s after it; here, the second RocksDB's share of the writes jumped from 0 to 52--54\%.
Any activity threshold up to half would have caught it.
After a profiling window of $4\hat A$ (356\,s), the new layout went in 417--420\,s after the switch in every run.
No write landed on another tenant's handle.

From then on, the layout derived by the loop alone, not offline as in Fig.~\ref{fig:stale}, kept WAF far below the static layout's for the whole hour.
Between 15 and 30\,min after the switch, WAF was 1.97 under the loop and 3.06 under the static layout in the three 30-min runs, and 1.97 and 3.04 in the 60-min run.
In the hour's last 15\,min, the loop was still 0.84 below, at 2.51 against 3.35, removing 36\% of the static layout's excess WAF.
Waiting for the profile cost no WAF later.
Installing the grouping at the switch did no better, at 2.40 between 15 and 30\,min and 2.70 in the last 15\,min.

\subsection{Does \tbdfixed{} Need Tuning?}
\label{sec:eval-sens}

\tbdfixed{} needed no per-workload tuning in our tests, because only two inputs shaped its layouts: the stream budget $B$ and the reclaim age $\hat A$.
Both are measured.
When we re-planned the profiles of one RocksDB and of the shared node one parameter at a time, its other parameters (Table~\ref{tab:params}) left the layout unchanged for any minimum side size from 1 to 20\% and any file-pattern purity from 60 to 100\%.
So did scoring survival anywhere from 0.40 to 1.2\,$\hat A$ on the shared node and from 0.13 to 2.6\,$\hat A$ on one RocksDB.
Past 1.2\,$\hat A$, generation\,4, which dies near $\hat A$, changes side, yet WAF stayed within 0.1.

The profiling window of four reclaim ages is long enough.
On the shared node, windows longer than 3.75\,$\hat A$ moved no further bytes between streams.
On one RocksDB, the first snapshot, at 0.45\,$\hat A$, was already final.
Thus, the earlier 900\,s window behind the shared-node layouts changed nothing, nor did their directory walk.
Recording paths at open instead changes no byte's stream.

Finally, the \tbdfixedplanner{}'s choice holds against its nearest rivals.
Of the ten ways to cut RocksDB's six generations into three contiguous classes, its $\{0\text{--}3\}\{4\}\{5\}$ matched the two it ranks next, $\{0\text{--}2\}\{3\text{--}4\}\{5\}$ and $\{0\text{--}1\}\{2\text{--}4\}\{5\}$, within 0.1 WAF.
The median-age ranking's $\{0\text{--}2\}\{3\}\{4\text{--}5\}$ was 0.31 worse.

\subsection{What Does Per-Write Placement Cost?}
\label{sec:eval-overhead}

Placement costs 8--9\% more CPU per small write, direct or buffered, against the same kernel without eBPF (Table~\ref{tab:overhead}).
The cost is fixed per request.
A 4\,KiB direct write takes about 4.5\,\textmu s more than its 56\,\textmu s of CPU, however little data it carries.
eBPF programs account for 2.6--3.0\,\textmu s of that (the placer 1.45--1.74), and the rest is unattributed.
These costs include the \tbdfixedmonitor{}'s signals.
Profiling adds 0.5--1.4\,\textmu s of eBPF time per small write and 108\,MiB of tables on the emulated SSD, but only during a profiling window.

Placement's CPU slows a writer only when it is CPU-bound.
At queue depth 16 and in buffered writes, where fio's thread was nearly saturated, IOPS fell by about as much.
At queue depth 1, where the thread mostly waits on the device, IOPS and 99th-percentile latency stayed within $\pm$2.5\%.
The same held for 128\,KiB writes, which spread the same work over 32 times the data, and for reads, which leave the placer early.

\begin{table}[tp]
\caption{
Overhead of \tbdfixed{}'s eBPF programs on the emulated SSD.
Key: 95\% interval
within $\pm$2.5\% ($\approx$0), up to $x$ ($\le x$), or over 15\% wide (--).
}
\label{tab:overhead}
\centering
\scriptsize
\setlength{\tabcolsep}{3pt}
\begin{tabularx}{\columnwidth}{@{}>{\raggedright\arraybackslash}X ccc@{}}
\toprule
 & \multicolumn{3}{c}{Overhead vs.\ no eBPF (\%)} \\
\cmidrule(l){2-4}
Workload & Throughput loss & p99 increase & CPU per I/O \\
\midrule
4~KiB direct, QD1      & $\approx$0 & $\approx$0 & 8.0 \\
4~KiB direct, QD16     & 7.9 & 4.5 & 8.6 \\
4~KiB buffered         & 9.5 & -- & 8.7 \\
128~KiB direct, QD16   & $\approx$0 & $\approx$0 & $\le$5.5 \\
4~KiB direct read, QD1 & $\approx$0 & $\approx$0 & $\le$4.8 \\
\bottomrule
\addlinespace[2pt]
\multicolumn{4}{@{}p{\columnwidth}@{}}{CPU: the experiment's cgroups. One fio tenant writes unplaced under every arm, so the cost is eBPF work alone, including the generation tracker (0.1~\textmu s per placed request).} \\
\end{tabularx}
\vspace{-12pt}
\end{table}
\section{Related Work}
\label{sec:related}


\noindent
\textbf{Lifetime-aware placement.}
Lifetime-aware systems group data by expected death, but none divides a scarce handle budget among tenants.
On multi-stream SSDs~\cite{kang2014multistream}, FDP's predecessor, systems inferred lifetimes without application hints.
AutoStream~\cite{yang2017autostream} infers them from how sequentially, often, and recently an address is written.
PCStream~\cite{kim2019fully} does so from each write's program context and lifetimes that overwrites and discards reveal.
SepBIT~\cite{sepbit} infers them from invalidation times.
Later systems add host structure, such as file-system metadata in FStream~\cite{fstream}, object identity in FlashAlloc~\cite{flashalloc}, file paths and observed behavior through a user-space shim in Valet~\cite{valet}, and database semantics~\cite{howtowritessds}.

\vspace{2pt}
\noindent
\textbf{FDP placement policies.}
Existing FDP systems rely on application hints, operator-defined mappings, or fixed file-system policies (Table~\ref{tab:sota})~\cite{torfs-github,allison2025fdpcache,lee2025slimio,chen2024fdpfs,qi2026f2fsfdp}.
Even io.stream~\cite{son2026iostream}, which shows that grouping by lifetime matters more than stream count, assigns lifetimes from domain knowledge and leaves dividing a tenant to future work.


\vspace{2pt}
\noindent
\textbf{eBPF-based storage systems.}
Prior eBPF storage work speeds up I/O but does not place data.
ExtFUSE~\cite{extfuse}, BPF for Storage~\cite{bpfstorage}, XRP~\cite{zhong2022xrp}, and BPF-oF~\cite{bpfof} accelerate I/O or push computation closer to storage.
The potential for eBPF-based stream allocation was raised at the Linux LSFMM+BPF 2025 summit~\cite{edge2025lwnfdp}.
\tbdfixed{} realizes this direction through transparent, workload-adaptive allocation without modifying applications, file systems, or kernel source.

\section{Conclusion}
\label{sec:conclusion}

This work shows that effective FDP placement requires more than isolating tenants: writes must be grouped by lifetime within tenants, scarce placement streams must be allocated by their expected GC savings, and assignments must adapt as active tenants change. 
These findings challenge the reliance on static, application-defined placement policies. 
\tbdfixed{} demonstrates that such decisions can be made transparently from kernel-observed I/O behavior, without modifying applications, file systems, or kernel source. By making lifetime-aware placement automatic and adaptive, \tbdfixed{} enables practical FDP use in multi-tenant storage systems.


\bibliographystyle{IEEEtran}
\bibliography{refs}

\end{document}